\documentclass[english,a4paper,fleqn]{llncs}
\input{preamble}

\usepackage{todonotes}
\renewcommand{\todo}[1]{}  
\newcommand{\todoall}[1]{\todo[color=gray,size={\fontsize{8pt}{9pt}\selectfont}]{#1}}

\title{An Institutional Framework for Heterogeneous\\
Formal Development in UML}
\author{Alexander Knapp\inst{1} 
   \and Till Mossakowski\inst{2} 
   \and Markus Roggenbach\inst{3}} 
\institute{Universit\"at Augsburg, Germany 
      \and Otto-von-Guericke Universit\"at Magdeburg, Germany
      \and Swansea University, UK}

\begin{document}

\maketitle

\begin{abstract}
We present a framework for formal software development with UML. In
contrast to previous approaches that equip UML with a formal
semantics, we follow an institution based heterogeneous approach. This
can express suitable formal semantics of the different UML diagram
types directly, without the need to map everything to one specific
formalism (let it be first-order logic or graph grammars). We show how
different aspects of the formal development process can be coherently
formalised, ranging from requirements over design and Hoare-style
conditions on code to the implementation itself. The framework can be
used to verify consistency of different UML diagrams both horizontally
(e.g., consistency among various requirements) as well as vertically
(e.g., correctness of design or implementation w.r.t.\ the
requirements).
\end{abstract}

\section{Introduction}

Historically, industrial standards on software quality merely
mentioned formal methods for the sake of completeness. Nowadays, each
new (edition of a) standard brings formal methods more to the
fore. Thanks to this trend, current standards elaborate on formal
methods and often recommend their use as an important means to
establish a design's correctness and robustness. Recent examples
include the 2011 version of the CENELEC standard on Railway
applications, the 2011 ISO 26262 automotive standard, or the 2012
Formal Methods Supplement to the DO-178C standard for airborne
systems.

In industrial software design, the Unified Modeling Language (UML) is
the predominately used development mechanism. In aerospace industry,
e.g., the company AEC uses the UML to define the software architecture
of aeroplane engine controllers through various levels of abstraction
from a layered architecture overview to a detailed class, operation
and attribute definition of the software components. This model is
then used for code generation. Typically, the software components
developed are either reactive in nature or the components are
logic-based and/or stateful in nature, where notations such as UML
state diagrams are used to define the required behaviour. Similarly,
the UML is used in micro-controller development in the automotive
sector. An example out of the medical sector is the development of
ventricular assist devices, to name just a few uses of UML for the
development of critical systems.

The UML is an OMG standard~\cite{uml-2.4.1-superstructure}, which
describes a family of languages. It offers 14 types of diagrams of
both structural and behavioural nature. A typical development by AEC
easily involves eight different UML diagrams. The OMG specification
provides an informal semantics of nine sub-languages in isolation. The
languages are mostly linked through a common meta-model, i.e., through
abstract syntax only. This situation leads to a gap between standards'
recommendation to apply formal methods, and current industrial
practice, which by using the UML lacks the semantic foundations to
apply such methods.  One common approach to deal with this gap is to
define a comprehensive semantics for the UML using a system model,
e.g., \cite{broy-et-al:uml2:3:2009,broy-et-al:uml2:4:2009}. However,
such an approach is a thorny business, as every detail has to be
encoded into one, necessarily quite complex semantics. Furthermore,
UML's widespread adoption in industry is primarily due to its
versatility; it lends itself to variations of usage, and different
subsets of the UML tend to be used in different ways by different
companies, leading to company or domain-specific variations.


In this paper, we outline a competing approach by providing a
heterogeneous semantics. In this approach, we express the meaning of a
model in a sub-lan\-guage/di\-a\-gram directly in an appropriate
semantic domain and look for connections given by the abstract syntax
of the UML specification or which can be gleaned from its semantic
description. This separation between the meaning of the individual
diagrams and how they are related allows our approach to adopt to
different methodologies, for instance an object-oriented approach or a
component-based one.

\section{Methodology}

\begin{wrapfigure}[12]{R}{0.4\textwidth}
\centering
\vspace{-2.4em}
\includegraphics[trim=6 6 6 6,clip,scale=0.65]{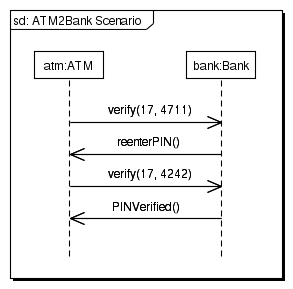}\\[-1ex]
\caption{\label{fig:interaction}Interaction.}
\end{wrapfigure}
Our overall aim is to providing Qualified Formal Methods for
dependable software design for critical systems, especially embedded,
reactive systems. This will cover requirements, design, code and
deployment.

\subsection{ATM case study}

In order to illustrate our heterogeneous semantics, we present as a
running example the design of a traditional automatic teller machine
(ATM) connected to a bank. For simplicity, we only describe the
handling of entering a card and a PIN with the ATM. After entering the
card, one has three trials for entering the correct PIN (which is
checked by the bank). After three unsuccessful trials the card is
kept.

\begin{wrapfigure}[7]{r}{0.55\textwidth}
\centering
\vspace{-2.4em}
\includegraphics[trim=6 6 6 6,clip,scale=0.65]{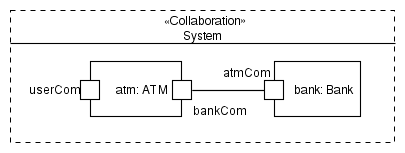}\\[-1.5ex]
\caption{\label{fig:system} Composite structure.}
\end{wrapfigure}

\paragraph{Requirements.}
Figure~\ref{fig:interaction} shows a possible \emph{interaction}
between an \uml{atm} and a \uml{bank} object, which consists out of
four messages: the \uml{atm} requests the \uml{bank} to \uml{verify}
if a card and PIN number combination is valid, in the first case the
\uml{bank} requests to reenter the PIN, in the second case the
verification is successful.
\begin{figure}[!Ht]
\centering
\includegraphics[trim=6 6 6 6,clip,scale=.65]{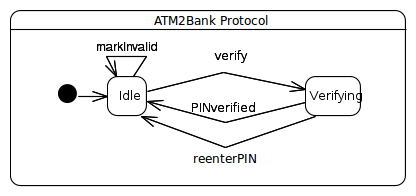}\\[-1.5ex]
\caption{\label{fig:psm}Protocol state machine.}
\end{figure}
This interaction presumes that the system has \uml{atm} and a
\uml{bank} objects. This can, e.g., be ensured by a \emph{composite
  structure diagram}, see Fig.~\ref{fig:system}, which --- among other
things --- specifies the objects in the initial system state.  In
order to communicate with a \uml{bank} object, we assume the \uml{atm}
object to have a \emph{behaviour port} called \uml{bankCom}. This port's
dynamic behaviour is captured with a protocol state machine, see
Fig.~\ref{fig:psm}.
\begin{figure}[!Ht]
\centering
\includegraphics[trim=6 6 6 6,clip,scale=.65]{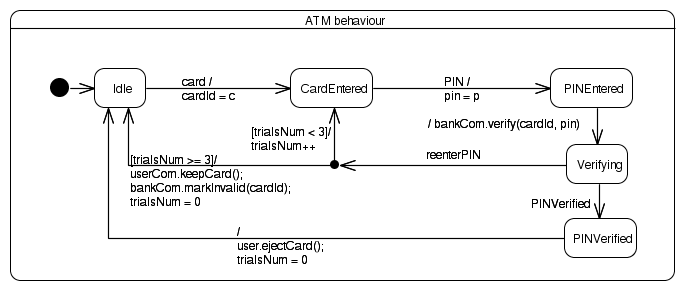}\\[-1.5ex]
\caption{\label{fig:state-machine}State machine.}
\end{figure}

\paragraph{Design.}
The dynamic behaviour of the \uml{atm} object is specified by the
\emph{state machine} shown in Fig.~\ref{fig:state-machine}. The
machine consists of five states including \uml{Idle},
\uml{CardEntered}, etc.  Beginning in the initial \uml{Idle} state,
the user can \emph{trigger} a state change by entering the
\uml{card}. This has the \emph{effect} that the parameter \uml{c} from
the \uml{card} event is assigned to the \uml{cardId} in the \uml{atm}
object (parameter names are not shown in a state machine
diagram). Entering a \uml{PIN} triggers another transition to
\uml{PINEntered}.  Then the ATM requests verification from the bank
using its \uml{bankCom} port.  The transition to \uml{Verifying} uses
a \emph{completion event}: No explicit trigger is declared and the
machine autonomously creates such an event whenever a state is
completed, i.e., all internal activities of the state are finished (in
our example there are no such activities).  In case the interaction
with the bank results in \uml{reenterPIN}, and the \emph{guard}
\uml{trialsNum < 3} is true, the user can again enter a \uml{PIN}.

\begin{wrapfigure}[5]{r}{0.4\textwidth}
\centering
\vspace{-2.4em}
\includegraphics[trim=6 6 6 6,clip,scale=0.65]{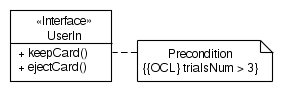}\\[-1.5ex]
\caption{\label{fig:class}Interface.}
\end{wrapfigure}
Figure~\ref{fig:class} provides structural information in form of an
interface diagram for the \uml{userCom} port of the \uml{atm}
object. An interface is a set of operations that other model elements
have to implement. In our case, the interface takes the form of a
\emph{class diagram}. Here, the operation \uml{keepCard} is enriched
with the OCL constraint \uml{trialsNum > 3}, which refines its
semantics: \uml{keepCard} can only be invoked if the OCL constraints
holds.

\paragraph{Code.} 
The state machine shown in Fig.~\ref{fig:state-machine} can be
implemented in the programming language C, enriched with
pre-/post-conditions written in the ANSI/ISO C Specification Language
(ACSL). The below code example shows how the event \uml{card} is
encoded as C function, where the ACSL annotations ensure that the
system is in some defined state and that the number of trials to
re-enter the PIN is smaller than three.
\begin{lstlisting}[basicstyle=\ttfamily\fontsize{8pt}{10pt}\selectfont,label={lst:intro}]
typedef enum states {
  EMPTY = 0,       IDLE = 1,       CARDENTERED = 2,
  PINENTERED = 3,  VERIFYING = 4,  PINVERIFIED = 5
} states_t;
int cardId = 0; int pin = 0; int trialsNum = 0;
states_t state = EMPTY;

/*@
  requires state != EMPTY; requires trialsNum <= 3;
  ensures state != EMPTY;  ensures trialsNum <= 3;
@*/
void card(int c) {
  switch (state) {
    case IDLE:
      cardId = c;
      state = CARDENTERED;
      break;
    default:
  }
}
...
\end{lstlisting}

\paragraph{Deployment.}
Finally, the \emph{composite structure diagram} in
Fig.~\ref{fig:system} shows the initial state of our ATM system, in
which an \uml{atm} object and a \uml{bank} object collaborate, where
the \uml{atm} has a \uml{bankCom} port, whilst the \uml{bank} has an
\uml{atmCom} port.

\subsection{From Requirements to Design to Code}

\begin{figure}[!Ht]
\centering
\includegraphics[scale=.21]{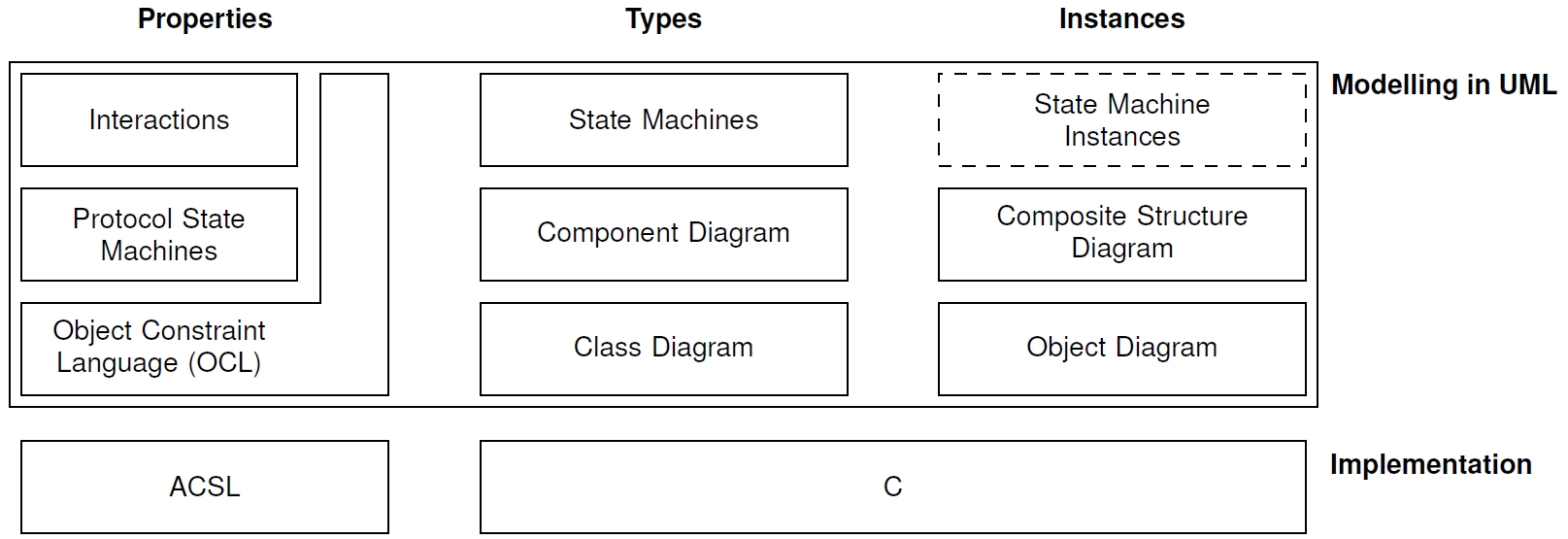}\\[-1.5ex]
\caption{Languages and diagrams considered}
\label{fig:languages}
\end{figure}

The languages and UML diagram types that we consider are shown in
Fig.~\ref{fig:languages}. On the \emph{modelling level} we use parts
of the UML and the Object Constraint Language (OCL). On the
\emph{implementation level} we currently employ the programming
language C and ACSL. It is left for future work to also include a
proper object-oriented language such as Java together with some
specification formalism. In the \emph{types view} of the modelling
level we look at class diagrams for modelling data; component diagrams
for modelling components; and state machines for specifying dynamic
behaviour.  These diagrams can be instantiated in the \emph{instance
  view} using composite structure diagrams for showing component
configurations; and object diagrams for showing concrete data.
Although they are not present in UML~2, we also have added state
machine instances (in a dashed box).  Requirements on the models can
be specified in the \emph{properties view} using interactions, i.e.,
sequence diagrams or communication diagrams, for prescribing message
exchanges between components and objects; protocol state machines for
specifying port behaviour; and the Object Constraint Language for
detailing the behaviour of components and objects in terms of
invariants and method pre-/post-conditions.

\subsection{Consistency and Satisfiability}

During a model-driven development, it is desirable to detect
inconsistencies at an early stage in order to ease corrections and
avoid costly re-engineering at a late stage (e.g.\ during the
implementation phase). While there are some tools providing static
inconsistency checks based on UML's meta-model, only few works
consider dynamic checks, and generally only for specific UML diagram
types, e.g. \cite{lano:uml2:2009}.

We will now outline a systematic method for the analysis of
UML models and their interrelation to code.  The analysis of UML
models can proceed either horizontally within the requirements or
within the design level checking for consistency within the level, or
vertically checking for satisfaction between these two levels. A
typical horizontal consistency check on the requirements level would
ask if the sequential composition of actions in an interaction diagram
is justified by an accompanying OCL specification. A typical vertical
satisfiability check between the requirements and the design level
would ask if the behaviour prescribed in an interaction diagram is
realisable by several state machines cooperating according to a
composite structure diagram. Code generation transforms a UML logical
design to code templates with semantic annotations in the form of
pre/post conditions and invariants. If the templates are completed
satisfying the semantic annotations, it is guaranteed that the
resulting code is a correct model of the logical design and therefore,
by the vertical checks, also for the requirements.

\section{UML as a Heterogeneous Formal Method, Using Institutions}

In this section, we will provide semantic foundations for model based
specification and design using a heterogeneous framework based on
Goguen's and Burstall's theory of institutions \cite{GoguenBurstall92}.  We
handle the complexity of giving a coherent semantics to UML by
providing several institutions formalising different diagrams of UML,
and several institution translations (formalised as so-called
institution morphisms and comorphisms) describing their interaction
and information flow. The central advantage of this approach over
previous approaches to formal semantics for UML
(e.g. \cite{lano:uml2:2009}) is that each UML diagram type can
stay ``as-is'', without the need of a coding using graph grammars (as
in \cite {DBLP:conf/uml/EngelsHK03}) or some logic (as in \cite{lano:uml2:2009}). This also keeps full flexibility in the choice of verification
mechanisms.
The formalisation of UML diagrams as institutions has the additional
benefit that a notion of refinement comes for free, see
\cite{caslref,CASL-refinement-journal}.

This systematic coverage in a single semantic based Meta-formalism is
unique.  We provide semantic links in the form of institution
(co-)morphisms, that, on the one hand, provide the basis for correct
model transformations and validations, and on the other hand give rise
to an integrated semantic view (via the so-called Grothendieck
institution \cite{Diaconescu02,Mossakowski02b}) on the identified UML subset as well as the
target implementation languages. Institution theory provides an
adequate abstraction level for such a semantic integration.
The framework is flexible enough to support various development
paradigms as well as different resolutions of UML's semantic variation
points.  This is the crucial advantage of the proposed approach to the
semantics of UML, compared to existing approaches in the literature
which map UML to a specific global semantic domain in a fixed way.

\subsection{Heterogeneous Formal Semantics of Languages and Diagrams}

To carry out this program of institutionalising UML in all detail goes
beyond scope and space limits of this paper. We only present some
cornerstones and sketch how this can be extended to all diagrams
in Fig.~\ref{fig:languages}.

For substantial fragments of several UML diagram types,
we have already provided a formalisation as institutions:
\begin{description}
\item[Class diagrams] in \cite{cengarle-et-al:ugo65:2008}, we
 have sketched an institution for class diagrams, which
has been detailed in \cite{DBLP:conf/wadt/JamesKMR12}. It includes
a construction for stereotypes.
\item[Component diagrams] form an institution similar to that for
  class diagrams.  The main difference are the connector types, which
  however are quite similar to associations.
\item[Object diagrams] are essentially reifications of
models of class diagrams.
\item[Composite structure diagrams] are similar to object
diagrams. The main difference are the connectors,
which however are quite similar to the links of object diagrams.
\item[Interactions] in \cite{cengarle-et-al:ugo65:2008}, we have
sketched an institution for interactions, as well as
their interconnection (also with class diagrams) via institution
comorphisms.
\item[OCL] in \cite{cengarle-et-al:ugo65:2008}, we have sketched
  institutions for OCL. In \cite{cengarle-knapp:sosym:2004}, the
  OCL semantics is presented in more detail. An institution based on
  this is in preparation.
\end{description}

Thus, the central remaining challenge for institutionalisting UML are
\textbf{state machines} and \textbf{protocol state machines}. Below,
we sketch institutions for these, which are very similar. Only their
sentences differ in that protocol state machines have a post condition
instead of an action. Post conditions can also speak about messages
being sent (using OCL).  \todo{Note that state machines are loose as
  well, namely in the effect of their actions, which need to be
  specified within a class diagram or dynamic logic.}
Formalising both \textbf{C} and \textbf{ACSL} as institutions is
future work. 

\subsection{Institutions and Their (Co)Morphisms}

Institutions are an abstract formalisation of the notion of logical
system. 
Informally, institutions provide four different logical notions:
sigantures, sentences, models and satisfaction.
Signatures provide the vocabulary that may appear in sentences
and that is interpreted in models.
The satisfaction relation determines whether a given sentence
is satisfied in a given model. The exact nature of signatures,
sentences and models is left unspecified, which leads to
a great flexibility. This is crucial for the possibility
to model UML diagrams (which in the first place are not
``logics'') as institutions.

An important feature of institutions is the presence of
signature morphisms, which cna be seen as mappings between
signatures. Sentences can be translated along signature
morphisms, and models reduced \emph{against} signature
morphisms. The satisfaction condition states that
satisfaction is invariant under change of notation
(along a signature morphism).

We briefly recall the formal definition of
institutions from~\cite{GoguenBurstall92}.
An institution $\institution{I} = (\instSig{I}, \instSen{I},
\instMod{I}, {\instmodels{I}})$ consists of (i)~a category of
\emph{signatures} $\instSig{I}$; (ii)~a \emph{sentence
  functor} $\instSen{I} : \instSig{I} \to
\category{Set}$, where $\category{Set}$ is the category of sets;
(iii)~a contra-variant \emph{model functor} $\instMod{I} :
(\instSig{I})\op \to \category{Class}$, where $\category{Class}$ is
the category of classes; and (iv)~a family of \emph{satisfaction
  relations} ${\instmodels[\Sigma]{I}} \subseteq
|\instMod{I}(\Sigma)| \times \instSen{I}(\Sigma)$ indexed
over $\Sigma \in |\instSig{I}|$, such that the following
\emph{satisfaction condition} holds for every signature morphism
$\sigma : \Sigma \to \Sigma'$ in $\instSig{I}$, every sentence
$\varphi \in \instSen{I}(\Sigma)$ and for every $\Sigma'$-model
$M' \in |\instMod{I}(\Sigma')|$:
\begin{equation*}
  \instMod{I}(\sigma)(M') \instmodels[\Sigma]{I} \varphi
\ \iff\ 
  M' \instmodels[\Sigma']{I} \instSen{I}(\sigma)(\varphi)
\ \text{.}
\end{equation*}
$\instMod{I}(\sigma)$ is called the \emph{reduct} functor (also
written ${-}\reductop\sigma$), $\instSen{I}(\sigma)$ the
\emph{translation} function (also written $\sigma({-})$).

It is possible to define standard logical notions like logical
consequence, logical theory, satisfiabilty etc.\ as well as
languages for structured specification and refinement in an
institution-independent way \cite{SannellaDT:FASFSD12}.


For relating institutions in a semantics preserving way, we consider
institution morphisms.  Given institutions
$\institution{I}$ and $\institution{J}$, an \emph{institution
  morphism} $\mu = (\Phi,\allowbreak \alpha,\allowbreak \beta) : \institution{I} \to
\institution{J}$ consists of (i)~a functor $\Phi : \instSig{I} \to
\instSig{J}$; (ii)~a natural transformation $\alpha : 
{\instSen{J}} \compfun \Phi \natto {\instSen{I}}$; and (iii)~a natural
transformation $\beta : \instMod{I} \natto
\instMod{J} \compfun \Phi\op$, such that the following \emph{satisfaction condition}
is satisfied for all $\Sigma \in |\instSig{I}|$, $M \in
|\instMod{J}(\Sigma)|$, and $\varphi' \in \instSen{I}(\Phi(\Sigma))$:
\begin{equation*}
  M \instmodels[\Sigma]{I} \alpha_{\Sigma}(\varphi')
\ \iff\ 
  \beta_\Sigma(M) \instmodels[\Phi(\Sigma)]{J} \varphi'
\ \text{.}
\end{equation*}

Dually, we consider
institution comorphisms.  Given institutions
$\institution{I}$ and $\institution{J}$, a \emph{simple institution
  comorphism} $\rho = (\Phi,\allowbreak \alpha,\allowbreak \beta) : \institution{I} \to
\institution{J}$ consists of (i)~a functor $\Phi : \instSig{I} \to
\instSig{J}$; (ii)~a natural transformation $\alpha : {\instSen{I}}
\natto {\instSen{J}} \compfun \Phi$; and (iii)~a natural
transformation $\beta : \instMod{J} \compfun \Phi\op \natto
\instMod{I}$, such that the following \emph{satisfaction condition}
is satisfied for all $\Sigma \in |\instSig{I}|$, $M' \in
|\instMod{J}(\Phi(\Sigma))|$, and $\varphi \in \instSen{I}(\Sigma)$:
\begin{equation*}
  M' \instmodels[\Phi(\Sigma)]{J} \alpha_{\Sigma}(\varphi)
\ \iff\ 
  \beta_\Sigma(M') \instmodels[\Sigma]{I} \varphi
\ \text{.}
\end{equation*}

The methodological need for these two kinds of mappings between
institutions will be explained in Sect. \ref{sec:trans} below.

\subsection{Towards an Institution for UML State Machines}
\label{sec:UML-SM}

\newcommand{\ENV}{\mathsf{Env}}
\newcommand{\SM}[1][]{\mathsf{SM}\text{$#1$}}

We will now formalise a simplified version of UML state machines as
institutions. In particular, we omit hierarchical states.  We start
with an institution for the \emph{environment} of a state machine.
This environment fixes the conditions which can be used in guards of
transitions, the actions for the effects of transitions, and also the
messages that can be sent from a state machine.  The source of this
information typically is a class or a component diagram: The
conditions and actions involve the properties available in the classes
or components, the messages are derived from the available signals and
operations.  The sentences of this environment institution form a
simple dynamic logic (inspired by OCL) which can express that if a
guard holds as pre-condition, when executing an action, a certain set
of messages has been sent out, and another guards holds as
post-condition.  We then build a family of institutions for
\emph{state machines} over this environment institution, which is
parameterised in the environment.  A state machine adds the events and
states that are used.  The events comprise the signals and operations
that can be accepted by the machine; some of these will, in general,
coincide with the messages from the environment.  Additionally, the
machine may react to completion events, i.e., internal events that are
generated when a state of the machine has been entered and which
trigger those transitions that do not show an explicit event as their
trigger in the diagrammatic representation (we use the states as the
names of these events).  The initial state as well as the transitions
of the machine are represented as sentences in the institution.  In a
next step, we combine the family of state machine institutions
parameterised over the environments into a single institution.
Finally, we present a product construction on the combined institution
that captures communicating state machines from a composite structure
diagram.

\paragraph{Environment institution.}
An object of the category of environment \emph{signatures}
$\instSig{\ENV}$ is a triple of sets
\begin{equation*}
  \Eta = (G_{\Eta}, A_{\Eta}, M_{\Eta})
\end{equation*}
of guards, actions, and messages; and a morphism $\Eta \to \Eta'$ of
$\instSig{\ENV}$ is a triple of functions $\eta : (\eta_G : G_{\Eta}
\to G_{\Eta'}, \eta_A : A_{\Eta} \to A_{\Eta'}, \eta_M : M_{\Eta} \to
M_{\Eta'})$.  The class of environment \emph{structures}
$\instMod{\ENV}(\Eta)$ for an environment signature $\Eta$ consists of
the triples
\begin{equation*}
  \Omega = (|\Omega|, \gamma_{\Omega} : G_{\Eta} \to \powerset |\Omega|, \alpha_{\Omega} : A_{\Eta} \to (|\Omega| \to |\Omega| \times \powerset(M_{\Eta})))
\ \text{,} 
\end{equation*}
where $|\Omega|$ is a set of data states, $\omega \in
\gamma_{\Omega}(g)$ expresses that the state $\omega \in |\Omega|$
satisfies guard $g$, and $(\omega', \overline{m}) =
\alpha_{\Omega}(a)(\omega)$ that action $a$ leads from state $\omega
\in |\Omega|$ to state $\omega' \in |\Omega|$ producing the set of
messages $\overline{m} \subseteq M_{\Eta}$.  The \emph{reduct}
$\Omega'\reductop\eta$ of an $\Eta'$-environment structure $\Omega'$
along the morphism $\eta : \Eta \to \Eta'$ is given by $(|\Omega'|,
\gamma_{\Omega'}\reductop\eta, \alpha_{\Omega'}\reductop\eta)$ where
$(\gamma_{\Omega'}\reductop\eta)(g) = \gamma_{\Omega'}(\eta_G(g))$ and
$(\alpha_{\Omega'}\reductop\eta)(a)(\omega') = (\omega'',
\eta_M^{-1}(\overline{m}'))$ if, and only if
$\alpha_{\Omega'}(\eta_A(a))(\omega') = (\omega'', \overline{m}')$.
The set of environment \emph{sentences} $\instSen{\ENV}(\Eta)$ for an
environment signature $\Eta$ comprises the expressions
\begin{equation*}
  g_{\mathrm{pre}} \rightarrow [a]\overline{m} \rhd g_{\mathrm{post}}
\end{equation*}
with $g_{\mathrm{pre}},\allowbreak g_{\mathrm{post}} \in G_{\Eta}$, $a
\in A_{\Eta}$, and $\overline{m} \subseteq M_{\Eta}$, intuitively
meaning (like an OCL constraint) that if the pre-condition
$g_{\mathrm{pre}}$ currently holds, then, after executing $a$, the
messages $\overline{m}$ are produced and the post-condition
$g_{\mathrm{post}}$ holds.  The \emph{translation}
$\eta(g_{\mathrm{pre}} \rightarrow [a]\overline{m} \rhd
g_{\mathrm{post}})$ of a sentence $g_{\mathrm{pre}} \rightarrow
[a]\overline{m} \rhd g_{\mathrm{post}}$ along the signature morphism
$\eta : \Eta \to \Eta'$ is given by $\eta_G(g_{\mathrm{pre}})
\rightarrow [\eta_A(a)]\powerset\eta_M(\overline{m}) \rhd
\eta_G(g_{\mathrm{post}})$.  Finally, the satisfaction relation
$\Omega \instmodels[\Eta]{\ENV} g_{\mathrm{pre}} \rightarrow
[a]\overline{m} \rhd g_{\mathrm{post}}$ holds if, and only if, for all
$\omega \in |\Omega|$, if $\omega \in
\gamma_{\Omega}(g_{\mathrm{pre}})$ and $(\omega', \overline{m}') =
\alpha_{\Omega}(a)(\omega)$, then $\omega' \in
\gamma_{\Omega}(g_{\mathrm{post}})$ and $\overline{m} \subseteq
\overline{m}'$.  Then the \emph{satisfaction condition} is shown
easily.

\begin{example}
Consider the UML component \uml{ATM} with its properties \uml{cardId},
\uml{pin}, and \uml{trialsNum}, its ports \uml{userCom} and
\uml{bankCom}, and its outgoing operations \uml{ejectCard()} and
\uml{keepCard()} to \uml{userCom}, and \uml{verify()} and
\uml{markInvalid()} to \uml{bankCom}.  An environment signature for
\uml{ATM} is derived by forming guards, actions, and messages over
this information, such that it will contain the guards \uml{true} and \uml{trialsNum
  == 0}, the actions \uml{user.ejectCard(); trialsNum = 0} and
\uml{trialsNum++}, as well as the messages \uml{user.ejectCard()} and
\uml{bank.markInvalid(cardId)}.  Environment sentence over such an
environment signature could be (for $n \in \NZ$)
\begin{gather*}
  \uml{true} \rightarrow [\uml{user.ejectCard(); trialsNum = 0}]\{ \uml{user.ejectCard()} \} \rhd \uml{trialsNum == 0}
\quad\text{or}\\
  \uml{trialsNum == $n$} \rightarrow [\uml{trialsNum++}]\emptyset \rhd \uml{trialsNum == $n+1$}
\ \text{.}
\end{gather*}
\end{example}

\paragraph{State machine institution.}
The institution of state machines is now built over the environment
institution.  Let $\Eta$ be an environment signature and $\Omega$ an
environment structure over $\Eta$.  An object of the category of state
machine \emph{signatures} $\instSig{\SM[(\Eta, \Omega)]}$ over $\Eta$
and $\Omega$ is given by the pairs
\begin{equation*}
  \Sigma = (E_{\Sigma}, S_{\Sigma})
\end{equation*}
of events and states with $E_{\Sigma} \cap S_{\Sigma} = \emptyset$;
and a morphism $\sigma : \Sigma \to \Sigma'$ of $\instSig{\SM[(\Eta,
  \Omega)]}$ is a pair of injective functions $\sigma = (\sigma_E :
E_{\Sigma} \to E_{\Sigma'}, \sigma_S : S_{\Sigma} \to S_{\Sigma'})$.
The class of state machine \emph{structures} $\instMod{\SM[(\Eta,
  \Omega)]}(E, S)$ for a state machine signature $(E, S)$ over $\Eta$
and $\Omega$ consists of the pairs
\begin{equation*}
  \Theta = (I_{\Theta}, \Delta_{\Theta})
\end{equation*}
where $I_{\Theta} \in \powerset |\Omega| \times S_{\Sigma}$ represents
the initial configurations, fixing the initial control state; and
$\Delta_{\Theta} \subseteq C \times \powerset(M_{\Eta}) \times C$ with
$C = |\Omega| \times \powerset(E_\Sigma \cup S_{\Sigma}) \times
S_{\Sigma}$ represents a transition relation from a configuration,
consisting of an environment state, an event pool, and a control
state, to a configuration, emitting a set of messages.  The event pool
may contain both events declared in the signature (from signals and
operations) and completion events (represented by states).  The
\emph{reduct} $\Theta'\reductop\sigma$ of a state machine structure
$\Theta'$ along the morphism $\sigma : \Sigma \to \Sigma'$ is given by
the structure $(\{ (\omega, s) \mid (\omega, \sigma_S(s)) \in I' \},
\{ (c_1, \overline{m}, c_2) \mid\allowbreak (\sigma_C(c_1),
\overline{m}, \sigma_C(c_2)) \in \Delta' \})$ where $\sigma_C(\omega,
\overline{p}, s) = (\omega, \powerset\sigma_P(\overline{p}),
\sigma_S(s))$ and $\sigma_P(p) = \sigma_E(p)$ if $p \in E_{\Sigma}$ and $\sigma_P(p) = \sigma_S(p)$
if $p \in S_{\Sigma}$.  The set of state machine \emph{sentences}
$\instSen{\SM[(\Eta, \Omega)]}(\Sigma)$ for a state machine signature
$\Sigma$ over $\Eta$ and $\Omega$ consists of the pairs
\begin{equation*}
  (s_0 \in S_{\Sigma}, T \subseteq S_{\Sigma} \times (E_{\Sigma} \cup S_{\Sigma}) \times G_{\Eta} \times A_{\Eta} \times S_{\Sigma})
\end{equation*}
where $s_0$ means an initial state and $T$ represents the transitions
from a state with a triggering event (either a declared event or a completion event), a guard, and an action to
another state.  The translation $\sigma(s_0, T)$ of a sentence $(s_0,
T)$ along the signature morphism $\sigma : \Sigma \to \Sigma'$ is
given by $(\sigma_S(s_0),\allowbreak \{ (\sigma_S(s_1), \sigma_P(p), g, a,
\sigma_S(s_2)) \mid (s_1, p, g, a, s_2) \in T \})$.  Finally, the
\emph{satisfaction relation} $\Theta \instmodels[\Sigma]{\SM[(\Eta,
  \Omega)]} (s_0, T)$ holds if, and only if $\pi_2(I_{\Theta}) = s_0$
and
\begin{gather*}
  \Delta_{\Theta} = \{ ((\omega, p \uplus \overline{p}, s), \overline{m} \cap (M_H \setminus E_{\Sigma}), (\omega', \overline{p} \lhd ((\overline{m} \cap E_{\Sigma}) \cup \{ s' \}), s')) \mid{}\\
\qquad\qquad\omega \in \gamma_{\Omega}(g),\ (\omega', \overline{m}) = \alpha_{\Omega}(a)(\omega),\ (s, p, g, a, s') \in T \}
\cup{}\\
  \phantom{\Delta_{\Theta} ={}}\{ ((\omega, p \lhd \overline{p}, s), \emptyset, (\omega, \overline{p}, s)) \mid \forall (s, p', g, a, s') \in T \,.\, p \neq p' \lor \omega \notin \gamma_{\Omega}(g) \}
\end{gather*}
where $p \uplus \overline{p}$ expresses that $p$ is the next event to
be processed by the machine according to some selection scheme from
the pool (where completion events are prioritized), and $\overline{p}
\lhd \overline{p}'$ adds the events in $\overline{p}'$ to the pool
$\overline{p}$.  The messages on a transition in the structure
$\Theta$ are only thosed that are not accepted by the machine itself,
i.e., not in $E_{\Sigma}$.  The accepted events in $E_{\Sigma}$ as
well as the completion event when entering state $s'$ are added to the
event pool of the target configuration.  When no transition is
triggered by the current event, the event is discarded (this will
happen, in particular, to all superfluously generated completion
events).  With these definitions, checking the satisfaction condition
\begin{equation*}
  \Theta'\reductop\sigma \instmodels[\Sigma]{\SM[(\Eta, \Omega)]} (s_0, T)
\iff
  \Theta \instmodels[\Sigma']{\SM[(\Eta, \Omega)]} \sigma(s_0, T)
\end{equation*}
for a state machine signature morphism $\sigma : \Sigma \to \Sigma'$
is straightforward.

\begin{example}
Consider the state machine of Fig.~\ref{fig:state-machine} defining
the behaviour of \uml{ATM}.  It works over the environment signature
sketched in the previous example, and its signature is
$(E_{\uml{ATM}}, S_{\uml{ATM}})$ with
\begin{gather*}
  E_{\uml{ATM}} = \{ \uml{card}, \uml{PIN}, \uml{reenterPIN}, \uml{PINVerified} \} \cup S_{\uml{ATM}}
\ \text{,}\\
  S_{\uml{ATM}} = \{ \uml{Idle}, \uml{CardEntered}, \uml{PINEntered}, \uml{Verifying}, \uml{PINVerified} \}
\ \text{.}
\end{gather*}
The state machine can be represented as the following sentence over this signature:
\begin{gather*}
  (\uml{Idle},
   \{ (\uml{Idle}, \uml{card}, \uml{true}, \uml{cardId = c}, \uml{CardEntered}),
\\\phantom{(\uml{Idle}, \{}
      (\uml{CardEntered}, \uml{PIN}, \uml{true}, \uml{pin = p}, \uml{PINEntered}),
\\\phantom{(\uml{Idle}, \{}
      (\uml{PINEntered}, \uml{PINEntered}, \uml{true}, \uml{bank.verify(cardId, pin)}, \uml{Verifying}),
\\\phantom{(\uml{Idle}, \{}
      (\uml{Verifying}, \uml{reenterPIN}, \uml{trialsNum < 2}, \uml{trialsNum++}, \uml{CardEntered}),
   \ldots \})
\ \text{.}
\end{gather*}
In particular, \uml{PINEntered} occurs both as a state and as a
completion event in the third transition.  The junction pseudostate
for making the decision whether \uml{trialsNum < 2} or \uml{trialsNum
  >= 2} has been resolved by combining the transitions.
\end{example}

\paragraph{Protocol state machine institution.}
Protocol state machines differ from behavioural state machines by not
mandating a specific behaviour but just monitoring behaviour: They do
not show guards and effects, but a pre- and a postcondition for the
trigger of a transition.  Moreover, protocol state machines do not
just discard an event that currently does not fire a transition; it is
an error when such an event occurs.

For adapting the state machine institution to protocol state machines
we thus change the \emph{sentences} to
\begin{equation*}
  (s_0 \in S_{\Sigma}, T \subseteq S_{\Sigma} \times G_{\Eta} \times E_{\Sigma} \times G_{\Eta} \times \powerset(M_{\Eta}) \times S_{\Sigma})
\end{equation*}
where the two occurrences of $G_{\Eta}$ represent the pre- and the
post-conditions, and $\powerset(M_{\Eta})$ represents the messages
that have to be sent out in executing the triggering event (protocol
state machines typically do not show completion events).  The
\emph{satisfaction relation} now requires that when an event $e$ is
chosen from the event pool the pre-condition of some transition holds
in the source configuration, its post-condition holds in the target
configuration, and that all messages have been sent out.  Instead of
the second clause of $\Delta_{\Theta}$, discarding an event, a
dedicated error state is targeted when no transition is enabled. %
\todoall{We could claim that there is a co-morphism from the
  protocol state machine to the environment institution and/or the OCL
  institution --- but this seems to be quite bold.}

\paragraph{Flat state machine institution.}
We now flatten the institutions $\SM[(\Eta, \Omega)]$ for each
environment signature $\Eta$ and each environment structure $\Omega$
over $\Eta$ into a single institution $\SM$.\footnote {This is an
  instance of a general construction, namely the Grothendieck
  institution \cite{Diaconescu02}.}  The signatures $\langle\Eta,
\Sigma\rangle$ consist of an environment signature $\Eta$ and a state
machine signature $\Sigma$, similarly for signature morphisms as well
as for structures $\langle\Omega, \Theta\rangle$.  As $\langle\Eta,
\Sigma\rangle$-sentences we now have both dynamic logic formulas (over
$\Eta$), as well as control transition relations (over $\Eta$ and
$\Sigma$).  Also satisfaction is inherited.  Only the definition of
reducts is new, because they need to reduce state machine structures
along more complex signature morphisms: $\langle\Omega',
\Theta'\rangle\reductop(\eta, \sigma) = \langle\Omega'\reductop\eta,
\Theta'\reductop\sigma\reductop\eta\rangle$ where
$\Theta''\reductop\eta = (\{ c''_1, \eta_M^{-1}(\overline{m}''),
c''_2) \mid (c''_1, \overline{m}'', c''_2) \in \Delta_{\Theta''} \},
I_{\Theta''})$.

Inside the flat state machine institution $\SM$ we can consider the
composition of state machines over different environments.  These
different environments represent the local views of the state
machines.  Given two state machine signatures $\langle\Eta_1,
\Sigma_1\rangle$ and $\langle\Eta_2, \Sigma_2\rangle$ of $\SM$ with
$G_{\Eta_1} \cap G_{\Eta_2} = \emptyset$, $A_{\Eta_1} \cap
A_{\Eta_2}$, $E_{\Sigma_1} \cap E_{\Sigma_2} = \emptyset$, and
$S_{\Sigma_1} \cap S_{\Sigma_2} = \emptyset$, we combine these into a
single signature $\langle\hat{\Eta}, \hat{\Sigma}\rangle$ of $\SM$ by
taking the component-wise union for the guard, actions, and messages,
the union of events and states for the events, and the product for the
states component.\footnote{The sharing of guards and actions could
  also be covered by a push-out construction.} Now, consider two state
machine structures $(\Omega_1, \Theta_1)$ over $\langle\Eta_1,
\Sigma_1\rangle$ and $(\Omega_2, \Theta_2)$ over $\langle\Eta_2,
\Sigma_2\rangle$, respectively.  Their \emph{interleaving product} is
given by
\begin{equation*}
  \langle\Omega_1, \Theta_1\rangle \pll \langle\Omega_2,
  \Theta_2\rangle = (\Omega_1 \pll \Omega_2, \Theta_1 \pll \Theta_2)
\end{equation*}
\begin{itemize}[label={--},topsep=2pt,itemsep=0pt,leftmargin=*]
  \item $\Omega_1 \pll \Omega_2 = (|\Omega_1| \times |\Omega_2|,
\gamma_{\Omega_1} \pll \gamma_{\Omega_2}, \alpha_{\Omega_1} \pll
\alpha_{\Omega_2})$ where $(\omega_1, \omega_2) \in (\gamma_{\Omega_1}
\pll \gamma_{\Omega_2})(\hat{g})$ if $\hat{g} \in G_{\Eta_i}$ and
$\omega_i \in \gamma_{\Theta_i}(\hat{g})$ for $i \in \{ 1, 2 \}$; and
$(\alpha_{\Omega_1} \pll \alpha_{\Omega_2})(\hat{a})(\omega_1,
\omega_2) = ((\omega_1', \omega_2'), \overline{\hat{m}})$ if $\hat{a}
\in A_{\Eta_i}$ and $\alpha_{\Omega_i}(\hat{a})(\omega_i) =
(\omega_i', \overline{\hat{m}})$ and $\omega'_j = \omega_j$ for $i
\neq j \in \{ 1, 2 \}$.

  \item $\Theta_1 \pll \Theta_2 = (I_{\Theta_1} \pll I_{\Theta_2}, \Delta_{\Theta_1} \pll \Delta_{\Theta_2})$ with $I_{\Theta_1} \pll
I_{\Theta_2} = ((s_1, s_2), \Omega_1 \times \Omega_2)$ for
$I_{\Theta_i} = (s_i, \Omega_i)$, and
\begin{gather*}
  \Delta_{\Theta} \pll \Delta_{\Theta'}
=
  \{ ((\omega_1, \omega_2), e \uplus (\overline{e}_1 \cup \overline{e}_2), (s_1, s_2)),
     \overline{m} \cap (M_{\hat{\Eta}} \setminus E_{\hat{\Sigma}}),
\\\phantom{\Delta_{\Theta} \pll \Delta_{\Theta'} = \{}
     ((\omega_1', \omega_2'), (\overline{e}_1 \cup \overline{e}_2) \lhd (\overline{e}' \cap E_{\hat{\Sigma}}), (s_1', s_2'))
\mid{}
\\\qquad\qquad\qquad\qquad
  \exists i \in \{ 1, 2 \} \,.\, ((\omega_i, e \uplus \overline{e}_i, s_i), \overline{m}, (\omega_i', \overline{e}_i \lhd \overline{e}', s_i')) \in \Delta_{\Theta_i}
\}
\ \text{.}
\end{gather*}
\end{itemize}

\begin{example}
Consider the composite structure diagram in Fig.~\ref{fig:system},
showing instances \uml{atm} and \uml{bank} of the \uml{ATM} and
\uml{Bank} components, respectively, that are connected through their
\uml{bankCom} and \uml{atmCom} ports.  In execution, \uml{atm} and
\uml{bank} will exchange messages, as prescribed by their state
machines, and this exchange is reflected by the interleaving product
which internalises those events that are part of the common signature.
On the other hand, messages to the outside, i.e., through the
\uml{userCom} port are still visible.
\end{example}

\subsection{Transformations Among UML institutions}
\label{sec:trans}

\begin{figure}[!Ht]
\centering
\includegraphics[scale=.21]{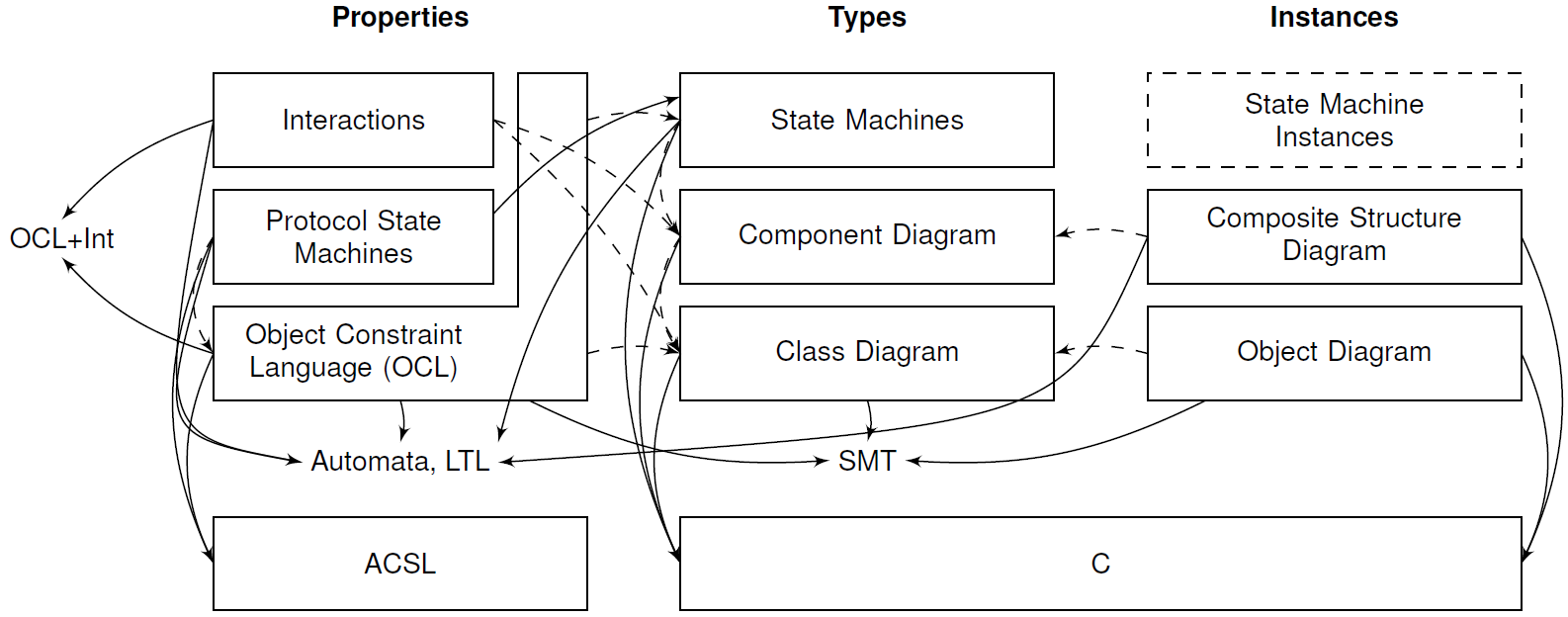}\\[-1.5ex]
\caption{Institution morphisms (dashed arrows) and institution
  co-morphisms (solid arrows) between the languages and diagrams}
\label{fig:language-relationships}
\end{figure}

Figure~\ref{fig:language-relationships} gives an overview of the
transformations between the modeling languages, diagram types, and
additional languages. The transformations in the figure can be
formalised as institution morphisms and comorphisms.  An institution
morphism (represented by a solid line in the figure) roughly
corresponds to a projection from a ``richer'' to a ``poorer'' logic,
expressing that the ``richer'' logic has some more features, which are
forgotten by the morphism. The main purpose of the institution
morphisms is the ability to express, e.g., that an interaction diagram
and a state machine are compatible because they are expressed over the
same class diagram.  Institution morphisms thus enable the
formalisation of heterogeneous UML specifications as structurded
specifications over the Grothendieck institution, a flattening of the
diagram of institutions and morphisms \cite{Diaconescu02}.
Practically, these structured Grothendieck specifications can be
formulated in the distributed ontology, modeling and specification
lanaguage (DOL), which currently is being standardized in the OMG (see 
\url{ontoiop.org} and~\cite{womo13-dol}).
 
By contrast, institution comorphisms (represented by dashed lines in
the figure) are often more complex. Roughly, a comorphism corresponds
to an encoding of one logic into a another one. The purpose of
institution comorphisms is threefold: \textbf{(1)} to provide a means
for expressing the dynamic checks (see below) in the institutional
framework, \textbf{(2)} to obtain tool support for the various UML
diagrams by using comorphisms into tool-supported institutions, and
\textbf{(3)} to transform UML diagrams into ACSL specifications and C
programs.

Dynamic checks and tool support involve additional institutions (also
depicted in Fig.~\ref{fig:language-relationships}, but not formalised
in detail here) for certain automata, like those used in the model
checker SPIN, and satisfiability modulo theories (SMT) provers, as
well as linear temporal logic.  The modeling language institutions can
be embedded into these, paving the way for tool and prover support.

\subsection{Consistency and Satisfiability, Revisited}

The horizontal dimension of the relationship between the different
models has to ensure \emph{consistency} of the
models, i.e., that the models fit together and describe a coherent
system.  The same has to be checked on the implementation level for
the consistency between the C program and the ACSL specification;
however, here we can reuse existing theory and tools.

There are different kinds of consistency checks on the modelling
level: Static checks ensuring type consistency and type correctness
between types and instances.  Dynamic checks include the properties
and one or several cooperating instances or types.  Most of the
dynamic checks are theoretically undecidable, thus fully automatic
tools will not be able to answer all instances. However, in many
cases, useful automatic approximations are possible, while in other
cases, manual effort may be involved.

\begin{figure}[!Ht]
\centering
\includegraphics[scale=.21]{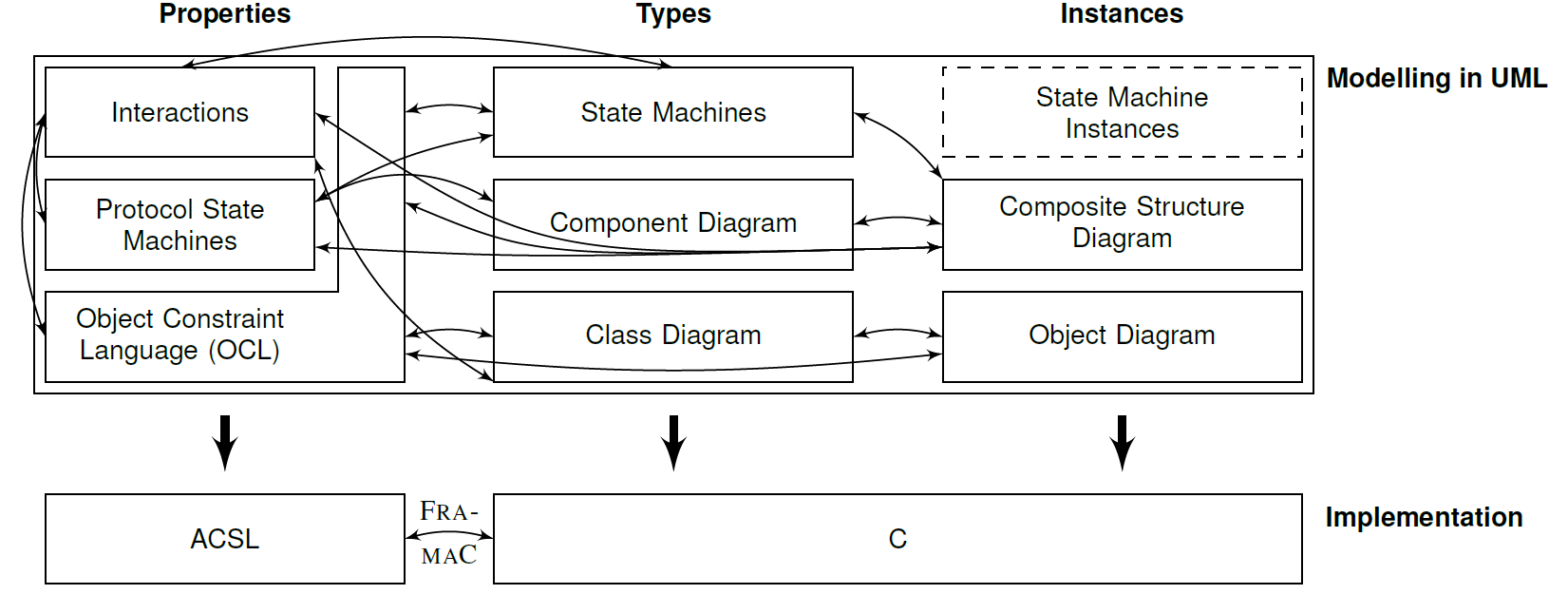}\\[-1.5ex]
\caption{Consistency relations (double-headed arrows) on the modelling and the implementation level; the bold
arrows represent the model transformations.}
\label{fig:consistency-relations}
\end{figure}
Figure~\ref{fig:consistency-relations} gives an overview of useful
relations between different kinds of diagrams, along which 
consistency checks are possible. We here only list a few of these.
Some useful static checks are:

\begin{enumerate}[topsep=0pt,itemsep=0pt,parsep=0pt,leftmargin=*]
  \item\label{it:ocl-comp-class} Does an OCL specification or a
composite structure diagram only use the methods of a class diagram?

  \item\label{it:sm-int-comp} Does a state machine or an interaction
comply with the interfaces in a composite structure diagram?

  \item\label{it:obj-class} Does an instance diagram (an object or a
composite structure diagram) comply to its corresponding type diagram
(a class or a component diagram)?

\end{enumerate}

Here are some useful dynamic checks:

\begin{enumerate}[topsep=0pt,itemsep=0pt,parsep=0pt,leftmargin=*]

  \item\label{it:sm-ocl} Does a state machine satisfy an OCL invariant
or an OCL pre-/post-condition?  

  \item\label{it:psms-comp} Do the protocol state machines at the ends
of a connector of a composite structure diagram fit together? 

  \item\label{it:int-sm-ocl} Is the sequential composition of methods
in an interaction diagram justified by the state machines and/or the
OCL specification?  


\end{enumerate}

For tackling question~\ref{it:sm-ocl}, we can use a semi-comorphism
from the OCL institution to the state machines institution that
selects those states and transitions that are relevant for the
invariant or the method with pre-/post-conditions.  However, the UML
does not specify the time point when the OCL post-condition should be
evaluated; one possibility is to choose the finishing of the fired
transition.

For question~\ref{it:psms-comp}, we can use a comorphism from the
protocol state machine institution into a temporal logic
institution~\cite{DBLP:books/daglib/0014211}, where we can form the
product of the protocol state machines (as detailed in
Sect.~\ref{sec:UML-SM}) along the connector.  However, the precise
nature of compatibility may be seen as a ``semantic variation point''.
Two important examples are the absence of deadlocks and buffer
overruns.

Concerning question~\ref{it:int-sm-ocl},
for the relation to an OCL specification we use a
co-span of institution comorphisms between the interactions
institution and the OCL institution
institution~\cite{cengarle-et-al:ugo65:2008}.  At least two links are
possible: In a strict interpretation, for each pair of successive
methods in the interaction there must be a state meeting the
post-condition of the first method and the pre-condition of the second
method.  In a more loose interpretation, a sequence of additional
method calls, not prescribed but also not excluded by the interaction,
must be possible to reach the pre-condition of the second method from
the post-condition of the first method.  For also considering state
machines, the co-span approach is extended by also involving the state
machines institution.

\section{Tools}

The Heterogeneous Tool Set (\protect\Hets)
\cite{MossakowskiEtAl06,MossakowskiEA06} provides analysis and proof
support for multi-logic specifications. The central idea of \Hets is
to provide a general framework for formal methods integration and
proof management that is equipped with a strong semantic
(institution-based) backbone. One can think of \Hets acting like a
motherboard where different expansion cards can be plugged in, the
expansion cards here being individual institutions (with their
analysis and proof tools) as well as institution (co)morphisms. The
\Hets motherboard already has plugged in a number of expansion cards
(e.g., SAT solvers, automated and interactive theorem provers, model
finders, model checkers, and more). Hence, a variety of tools is
available, without the need to hard-wire each tool to the logic at
hand. Via suitable translations, new formalisms can be connected to
existing tools.

We have just started to integrate first institutions for UML, such as
class diagrams, into \Hets. In order to obtain proof support for the
methodology presented in this paper, beyond the individual
institutions, also the morphisms and comorphisms need to be
implemented in \Hets. Moreover, we plan to connect \Hets to the tool
HugoRT \cite{knapp-merz-rauh:ftrtft:2002}. HugoRT can, on the one
hand, perform certain static checks on UML diagrams. Moreover, it
provides transformations of UML diagrams to automata and linear
temporal logic formulas, which can then be fed into model checkers
like SPIN in order to check certain properties.  The crucial benefit
of our approach is a clear separation of concerns: verification
conditions for consistency and satisfiability checks can be formulated
abstractly in terms of the UML institutions and (co)morphisms
described above. In a second step, these checks can then be
reformulated in terms of specific logics and tools that have been
connected to \Hets.

\section{Conclusion}

We have outlined an institution-based semantics for the main UML
diagrams, and in particular have provided an initial institution
for UML state machines as the main previously missing bit in the
overall picture. Moreover, we have sketched a methodology have
consistency among UML diagrams and with implementation languages
can be modeled at the institutional level and supported with tools.
 
Much remains to be done to fill in the details. The greatest missing
bit is certainly the institutional formalisation of programming
languages and their Hoare logics, like C and ACSL, or Java and
JML. Here, we want to follow the ideas sketched by A.~Tarlecki and
D.~Sannella~\cite[Ex.~4.1.32, Ex.~10.1.17]{SannellaDT:FASFSD12} for
rendering an imperative programming language as an institution.  The
semantic basis could be a simplified version of the operational
semantics of C.~Ellison and G.~Rosu~\cite{DBLP:conf/popl/EllisonR12}.
The concepts for institutionalising a Hoare logic like ACSL on the
basis of its specification~\cite{baudin-et-al:acsl-1.6:2012} can be
similar as for OCL.


\bibliographystyle{abbrv} 
\bibliography{uml,acsl,inst,hets,uml-inst}

\end{document}